%% file: main.tex
\documentclass[ALICE,manyauthors]{cernphprep}
\usepackage[comma,square,numbers,sort&compress]{natbib}
\usepackage{hyperref}
\usepackage{lineno}
\usepackage{xspace}
\usepackage{comment}
\usepackage{xcolor}
\usepackage[T1]{fontenc}
\usepackage{orcidlink}

\begin{document}
\input{commands.tex}

\begin{titlepage}
\PHyear{2025}       
\PHnumber{159}      
\PHdate{11 July}  

\title{Multiplicity dependence of K$\mathbf{^*(892)^{\pm}}$ production in pp collisions at $\mathbf{\sqrt{{s}}~=~13}$~\TeV}
\ShortTitle{Multiplicity dependence of K$^*(892)^{\pm}$}   

\Collaboration{ALICE Collaboration\thanks{See Appendix~\ref{app:collab} for the list of collaboration members}}
\ShortAuthor{ALICE Collaboration} 

\begin{abstract}
The first results of K$^*$(892)$^{\pm}$ production at midrapidity ($|y| < 0.5$) in pp collisions at \thirteen as a function of the event multiplicity are presented. The K$^*$(892)$^{\pm}$ has been reconstructed via its hadronic decay channel K$^*$(892)$^{\pm} \rightarrow~\pi^{\pm}~+~\kzero$~using the ALICE detector at the LHC. For each multiplicity class, the differential transverse momentum (\pt) spectrum, the mean transverse momentum \meanpt, the \pt-integrated yield (d$N$/d$y$), and the ratio of the K$^*$(892)$^{\pm}$ to \kzero yields are reported. These are consistent with previous K$^*$(892)$^0$ resonance results with a higher level of precision. Comparisons with phenomenological models such as PYTHIA6, PYTHIA8, EPOS-LHC, and DIPSY are also discussed. For the first time, a significant K$^*$(892)$^{\pm}$/\kzero suppression in pp collisions is observed at a 7$\sigma$~level passing from low to high multiplicity events. The ratios of the \pt-differential yields of K$^*$(892)$^{\pm}$ and  \kzero in high and low multiplicity events are also presented along with their double ratio.
For $\pt \lesssim 2$~\GeVc this double ratio persists below unity by more than $3\sigma$~suggesting that the suppression affects mainly low \pt resonances. The measured decreasing trend of the K$^*$(892)$^{\pm}$/\kzero ratio with increasing multiplicity, which in heavy-ion collisions is typically attributed to the rescattering of decay particles of the short-lived resonances, is reproduced by the EPOS-LHC model without the use of hadronic afterburners.

\end{abstract}
\end{titlepage}

\setcounter{page}{2} 


\section{Introduction}\label{sec:intro}
Recent studies on pp and p--Pb collisions at the LHC for events with high charged-particle multiplicities have shown patterns that are reminiscent of phenomena observed in heavy-ion collisions~\cite{alice3}. Those, for light flavor probes, include the strangeness enhancement~\cite{strange}, the hardening of hadron transverse momentum (\pt) spectra with increasing multiplicity~\cite{charged_hadr,KStar_pp7}, and the presence of long-range multiparticle azimuthal correlations~\cite{azimuthal,CMS:2016ppCollectivity,CMS:2016ppCorrelations,PHENIX:2015dAuCorrelations}. 
These phenomena are understood as due to the formation of the quark–gluon plasma (QGP)~\cite{alice3} in ultrarelativistic heavy-ion (A--A) collisions, where the pseudocritical values of temperature and energy density ($T_c \simeq 155$~\MeV and $\epsilon_c \simeq$~0.5-1~\GeV/fm$^3$, according to lattice-QCD calculations~\cite{Equation_state1,Equation_state2}) of the cross-over from the confined hadronic matter to the deconfined QGP state can be achieved. The QGP created in the collision expands and cools down until it reaches the temperature of the transition to a hadron gas. After the hadronization, the hadron gas evolves and cools down until the temperatures of the chemical and the kinetic freeze-out are reached. At the chemical freeze-out the abundances of the different hadron species are fixed (except for resonance decays). There are still significant elastic interactions between hadrons, affecting the momentum distributions of the different particle species that persist until the kinetic freeze-out occurs, when the elastic collisions also cease.

Traditionally, it was assumed that small collision systems, such as pp, evolve very differently from heavy-ion collisions. Hadronic collisions were thought to involve only multiparticle production in the final state, without any QCD phase transition. However, the unprecedented energies achieved at the LHC have challenged this view,  opening the possibility of the formation of QGP droplets in these collision systems.   This hypothesis continues to be an open question despite different recent measurements at RHIC ~\cite{PHENIX_NatPhys_2019,PHENIX_PRL_2018,PHENIX_PRC_2017,STAR_arXiv_2510_19645_2025,PHENIX_PRL_2013_dAu_quadrupole,STAR_PRL_130_242301_2023,STAR_PRC_110_064902_2024,STAR_PRL_122_172301_2019}  and LHC~\cite{ATLAS_PRL_2016,CMS_PLB_2017_pp_collectivity,CMS_PLB_2013_pPb_ridge,ALICE_PLB_2013_pPb_double_ridge,ATLAS_PRL_2013_pPb_ridge,ALICE_PRL_123_142301_2019}.

Short-lived hadronic resonances with a lifetime comparable to the duration of the hadronic phase, such as K$^*$(892) ($\tau$~$\approx$~4~fm/$c$), are perfect probes to characterize the evolution of the late hadronic phase, as they may be sensitive to the competing mechanisms of rescattering and regeneration~\cite{res_reg, res_reg2}. When a resonance decays inside the hadronic medium, the decay products may rescatter with the other particles of the hadron gas. Since resonances are reconstructed from the invariant mass distribution of their decay products, the rescattering effect causes a loss of reconstructed resonances, leading to a suppression of the measured resonance yield with respect to the abundance produced at the chemical freeze-out. However, resonances may also be regenerated by pseudo-elastic collisions of the particles of the hadronic medium, resulting in an enhancement of the measured yield. Longer-lived resonances, such as $\phi$(1020) ($\tau$~$\approx$~46~fm/$c$) decaying mainly at the end or after the hadronic phase, should not be affected by any of these processes. The best way to quantify the net effect for the suppression or enhancement of the reconstructed resonance yields is by comparing them to ground-state hadrons with similar quark content. The multiplicity-dependent suppression of ratios of \pt-integrated particle yields $\rho(770)^0/\pi$~\cite{rho,Adams2004rho}, K$^*(892)^0/$K~\cite{KStar5, KStarPbPb, KStar2, KStar2.1,Adler2002kstar,Adams2005kstar,Aggarwal2011kstar,Anticic2011kstar}, $\Lambda(1520)/\Lambda$~\cite{lambda,Abelev2006lambda}, and K$^*(892)^{\pm}/$K~\cite{KstarPMPbPb} were observed in heavy-ion collisions at ultrarelativistic energies, which suggests the dominance of rescattering over regeneration.
A first estimate of the lower limit for the duration of the hadronic phase was done  using the K$^*(892)^0/$K ratio measured in \pp, \pPb and heavy-ion collisions at RHIC and LHC energies~\cite{lifetime_RHIC}. From this analysis, it results that the hadronic phase lifetime strongly depends on the final-state charged-particle multiplicity, system size, and collision energy. In particular, the hadronic lifetime estimated for \pp collisions at LHC energies, 0--2~fm/$c$ for low and high multiplicity event is similar to the ones estimated in heavy-ion collisions at RHIC energies. Longer lifetimes of about 2--6~fm/$c$ are estimated for heavy-ion collisions at the LHC energies.

It is thus interesting to study the production of resonances as a function of multiplicity in \pp collisions to complement the results obtained in heavy-ion collisions and to provide possible further insight into the properties of the system created in small collision systems. A hint of suppression of the K$^*(892)^0/$K yield ratios was measured in high-multiplicity \pPb and \pp collisions~\cite{KStar_pPb, KStar_pp7, KStar_pp13} that could be interpreted as an evidence for the possible presence of a hadronic phase 
with a non-zero lifetime even in small collision systems. These results should be improved and corroborated by measurements with higher precision, as the K$^*(892)^0/$K suppression from low to high multiplicity events is measured with a significance of about 2$\sigma$~\cite{KStar_pPb, KStar_pp7, KStar_pp13}. 
It is possible to increase the precision by measuring charged instead of neutral K$^*(892)$ resonances, as observed in Ref.~\cite{incl_KStarpm}. The systematic uncertainties linked to the reconstruction of charged K$^*(892)$ are expected to be lower than those of their neutral counterparts in ALICE, because of the different strategy used for the K$^0_S$ and K$^{\pm}$ identification~\cite{KStar_pp7, sys_Ks}.

This article reports the first measurement of the multiplicity evolution of the K$^*$(892)$^{\pm}$ production at midrapidity ($|y|<$0.5) in pp collisions at the center-of-mass energy of \thirteen at the LHC with the ALICE detector. In particular, the \pt-differential yields, the mean transverse momenta $\langle \pt \rangle$, the \pt-integrated yields (d$N$/d$y$), as well as the ratios of the K$^*(892)^{\pm}$ to the \kzero yields, are studied as a function of the event multiplicity. Comparisons with the existing K$^*(892)^0$ measurements at the same collision energy~\cite{KStar_pp13} and model predictions are also presented. The results presented here achieve higher precisions than previous K$^*(892)^0$ measurements~\cite{KStar_pp13} and represent the first evidence of a significant suppression of the K$^*$(892)/K ratio in small collision systems.

The paper is organized as follows: the ALICE detector and the event selection criteria are described in Sec.~\ref{sec:alice}, with a specific focus on the main sub-detectors involved in performing the present analysis; Sec.~\ref{sec:analysis} provides details on the data analysis procedure with the evaluation of the systematic uncertainties; results and data comparisons are reported and discussed in Sec.~\ref{sec:results}, followed by summary and conclusion in Sec.~\ref{sec:conclusion}.

In the following, K$^{*0}$ is used to denote K$^*(892)^0$ and $\overline{\rm{K^*(892)^0}}$, K$^{*\pm}$ to denote K$^*(892)^{\pm}$, while K$^*$ indicates in general both K$^{*0}$ and  K$^{*\pm}$.

\section{Experimental setup and event selection}\label{sec:alice}
A detailed description of the ALICE apparatus and its performance can be found in Refs.~\cite{alice1, alice2}. The main sub-detectors used to accomplish the analysis reported in this paper are the Inner Tracking System (ITS)~\cite{alice1}, the Time Projection Chamber (TPC)~\cite{tpc}, and the V0A and V0C scintillators~\cite{vzero}. All tracking detectors are positioned in a solenoidal magnetic field of 0.5~T parallel to the LHC beam axis.

Charged particle tracks are reconstructed with the TPC and the ITS detectors, which provide high tracking efficiency down to \pt$\approx$ 0.1 \GeVc~\cite{alice2}. The ITS is the detector closest to the interaction point, directly surrounding the beam pipe and covering the region of radius between 4 and 43~cm and covering the pseudorapidity interval $|\eta|<$0.9. It is essential in the determination of the primary vertex and consists of six concentric cylindrical layers, based on three different types of silicon detectors: pixels (SPD), drifts (SDD), and strips (SSD), from the innermost to the outermost region, respectively. The ITS detectors have a spatial resolution of the order of a few tens of $\mu$m, with the best precision (12~$\mu$m) for the SPD detectors, which cover the pseudorapidity range $|\eta| < 1.3$. 

The TPC is the main tracking detector of ALICE, covering the pseudorapidity range $|\eta| < 0.9$. It has a cylindrical shape with an overall length along the beam direction of 500~cm and with inner and outer radii of $\simeq 90$~cm and $\simeq 250$~cm, respectively. Particle identification (PID) with the TPC is performed by means of the charge 
collected in the readout pads, which, in turn, is used to measure the specific ionization energy loss with a resolution of about 5--6\% at low $\pt$. A separation between $\pi$ and K and K and p larger than $\sim 2 \sigma$~is possible for $\pt < 0.8$~\GeVc and 1.6~\GeVc, respectively~\cite{alice2}.

The V0 detector consists of two arrays of scintillator tiles, called V0A and V0C, placed along the beam axis on each side of the nominal interaction point (IP) at $z~=~340$~cm  and $z~=~-90$~cm and covering the pseudorapidity regions $2.8 < \eta < 5.1$ (V0A) and $-3.7 < \eta < -1.7$ (V0C). With a time resolution better than $1$~ns, they are used for global event characterization, triggering, and beam background suppression.

The analysis presented in this paper is performed on the data sample of pp collisions at \thirteen collected with the ALICE apparatus during 2016, 2017, and 2018. The events were selected with a minimum bias trigger, which is given by the logical AND of the V0A and V0C detectors in coincidence with the arrival of proton bunches from both beam directions. A total number of about 1.3 $\times 10^{9}$ events was used for this analysis. The contamination from beam--gas interactions is removed offline using the timing information from the V0 detector~\cite{vzero}. Beam-induced background and pile-up events are also rejected by exploiting the correlation between the number of SPD hits and of SPD tracklets, which are track segments defined by pairs of clusters, one in each SPD layer, as explained in Ref.~\cite{alice2}. Selected events are required to have a primary collision vertex reconstructed from SPD tracklets and located along the beam axis within $\pm$10~cm from the center of the ALICE detector. For the multiplicity dependent analysis the \lq\lq INEL$>0$" class is used. It is defined as the set of inelastic collisions with at least one charged particle in the pseudorapidity range $|\eta| < 1$~\cite{inel} .
The INEL$>0$ sample is divided into several V0M multiplicity  classes based on the total charge deposited in both V0 detectors. The average charged-particle multiplicity densities at midrapidity \avdndetamid related to the different multiplicity classes are listed in Table~\ref{tab:mult_den}. They are indicated with Roman numbers, with the first class (I) corresponding to high multiplicity events, while the last one (X) refers to low multiplicity ones.

\begin{table}[h!]
\caption{Average charged-particle multiplicity densities at midrapidity (\avdndetamid) for various multiplicity classes in pp collisions at \thirteen~\cite{charged_hadr}}.
\label{tab:mult_den}
\centerline{
\begin{tabular}{|c|c|c|}
\hline 
Multiplicity (\%) & V0M Class & \avdndetamid\\
\hline 
-- & INEL $>$ 0 & 6.94 $\pm$ 0.10 \\
0--1 & I & 26.02 $\pm$ 0.35 \\
1--5 & II & 20.02 $\pm$ 0.27 \\
5--10 & III & 16.17 $\pm$ 0.22 \\
10--20 & IV+V & 12.91 $\pm$ 0.18 \\
20--30 & VI & 10.02 $\pm$ 0.14 \\
30--40 & VII & 7.95 $\pm$ 0.11 \\
40--50 & VIII & 6.32 $\pm$ 0.09 \\
50--70 & IX & 4.50 $\pm$ 0.07 \\
70--100 & X & 2.55 $\pm$ 0.04 \\
\hline 
\end{tabular}
}
\end{table}

It is worth to note as the average charged particle multiplicity densities at midrapidity related to the \pp collision at \thirteen, reported in Table 1,  for the multiplicity intervals 50--70\% and 0--1\% are similar to the ones for the intervals 80--100\% (peripheral) and 20\%-40\% (semicentral) for \pPb collisions at \fivenn~\cite{KStar_pPb}. Meanwhile, the \dndeta values for the interval 5-10\% are similar to the values for the peripheral collisions (80-90\%) of \PbPb collisions at \fivenn~\cite{ALICE2020_PRC101_044907}.

\section{Data analysis}\label{sec:analysis}

\subsection{The K$\mathbf{^{*\pm}}$ selection}\label{subsec:selection}
The K$^{*\pm}$, due to its short lifetime ($\tau \sim 10^{-23}$~s), cannot be directly measured by the ALICE detector. Therefore, it is reconstructed via its hadronic decay channel K$^{*\pm} \rightarrow~\pi^{\pm}~+~\kzero$ with a branching ratio (B.R.) of ($33.300 \pm 0.003$)\%~\cite{pdg2020}, which is the combined effect of the K$^{*\pm} \rightarrow~\pi^{\pm}~+~{\rm K}^{0}$ decay and of the probability of ${\rm K}^{0}$ to be in a \kzero state. The \kzero is reconstructed via its decay into an oppositely charged pion pair (V$^0$ topology), with a B.R. of ($69.2 \pm 0.05$)\%~\cite{pdg2020}. Therefore, the particles directly detected for the K$^{*\pm}$ reconstruction are charged pions. The pion from the K$^{*\pm}$ decay is denoted as a primary track because it originates from the interaction point, while the pions originating from the \kzero decay are secondary tracks. According to ALICE convention~\cite{Primary_track}, primary particles are those produced directly in the interaction together with all the decay products from particles with a mean proper decay length ($c\tau) < 1$~cm. All the other particles measured in the experiment are called instead secondary particles, including e.g., particles produced through interactions with the detector material and coming from weak decays. The distance of closest approach (DCA) of the track to the primary vertex is commonly used to discriminate primary and secondary particles. The reconstructed tracks are selected by applying a set of standard high-quality criteria using the information from the ITS and the TPC. The same selection criteria described in detail in Ref.~\cite{incl_KStarpm} are used to select primary and secondary tracks.  
Pions are identified through their specific ionization energy loss (d$E$/d$x$) measured in the TPC. The measured value of d$E$/d$x$ is required to lie within 3 standard deviations ($\sigma_{\rm TPC}$) from the specific energy loss expected for pions in the case of primary particles, while a looser selection of $5 \sigma_{\rm TPC}$~is required for secondary pions.

The V$^{0}$ reconstructed from the identified secondary tracks has to fulfill some topological criteria. The secondary vertex (i.e. the \kzero decay point) should have a radial distance from the primary vertex larger than 0.5~cm and the cosine of the pointing angle, i.e. the angle between the V$^0$~momentum and the vector defined by the primary and secondary vertices, is required to be larger than 0.97. 
The DCA between the V$^{0}$ daughter tracks is required to be smaller than one standard deviation with respect to the ideal value of zero, and the DCA of the V$^{0}$ to the primary vertex is required to be less than 0.3~cm. The difference between the invariant mass of the $\pi^+\pi^-$ pairs and the nominal \kzero mass ($m$~=~0.497611~\GeVmass\cite{pdg2020}) should be lower than 0.03~\GeVmass. This value is about 8.5(5) times the mass resolution value for \kzero with a transverse momentum of 0(10)~\GeVc. 
The \kzero candidates are further selected by requiring their rapidity to be in the range $|y|<0.8$ and their proper lifetime ($mL/p$) to be lower than 20~cm/$c$, where $L$ is the distance between the primary and secondary vertices and $p$ is the \kzero momentum. Finally, the V$^{0}$ candidates are rejected if their mass is compatible with the ${\Lambda}$ or 
${\overline{\Lambda}}$ mass [\cite{pdg2020}] within 0.0043~\GeVmass,
which is about three times the typical mass resolution for the reconstructed $\Lambda$ ($\overline{\Lambda}$) in ALICE~\cite{res_lam}. 
These selection criteria were varied from the default value in order to evaluate the systematic uncertainties, as explained in Sec.~\ref{subsec:sys}.

\subsection{Signal extraction}\label{subsec:signal}
The K$^{*\pm}$ raw yield in the rapidity range $|y|<0.5$ is extracted from the same-event invariant mass distribution of \kzero$\pi^{\pm}$  pairs in different \pt intervals in the range $0 < \pt < 10$~\GeVc and multiplicity classes (see Table~\ref{tab:mult_den}), similar to the procedure used for the K$^{*0}$ multiplicity dependent analysis performed by ALICE at the same collision energy and system~\cite{KStar_pp13}. 
The shape of the uncorrelated background is estimated via the event-mixing technique, by evaluating the invariant mass distribution of pairs taken from different events. To ensure a similar event structure and to avoid mismatch due to different acceptance, events are mixed only if they have similar vertex position along the $z$ direction and track multiplicity. To reduce statistical uncertainties, each event is mixed with nine other events. The mixed-event distribution is finally normalized to the same event distribution in the invariant mass region of $1.1 < {M_{\kzero\pipm}} < 1.2$~\GeVmass. The normalization range for the mixed-event background is one of the parameters varied for the study of the systematic uncertainties related to the signal extraction procedure.

After the combinatorial background subtraction, the remaining invariant mass distribution was fitted with a non-relativistic Breit--Wigner function to model the resonance peak and an ad-hoc function to describe the residual correlated background ($F_{BG}$):

\begin{equation}
 F(M_{\kzero\pipm}) = \frac{D}{2\pi} \frac{\Gamma_0}{(M_{\kzero\pipm} - M_0)^2 + \frac{\Gamma^2_0}{4}} + F_{BG}(M_{\kzero\pipm}){\color{red},}
\label{eq:eq1}
\end{equation}

where $D$ is the integral of the peak function from $0$ to $\infty$, while $M_0$ and $\Gamma_0$ are the mass and the width of K$^{*\pm}$. For the default measurement of the yields, the width was fixed to its natural value $\Gamma_0~=~(0.0508 \pm 0.0009)$~\GeVmass~\cite{pdg2020}, whereas for the estimation of the systematic uncertainties it was kept as a free parameter or fixed to different values ($\Gamma_1 = 0.0499$~\GeVmass and $\Gamma_2 = 0.0517$~\GeVmass).

As for the analysis of K$^{*\pm}$ production in inelastic pp collisions~\cite{incl_KStarpm}, the residual background shape is parameterized with the following function:

\begin{equation}
F_{BG}(M_{\kzero\pipm})  = [M_{\kzero\pipm} - (m_{\pipm} + m_{\kzero})]^n \, e^{(A + BM_{\kzero\pipm} + CM^2_{\kzero\pipm})},
\label{eq:eq2}
\end{equation}

where $m_{\pipm} = (0.13957061 \pm 0.00000002)$~\GeVmass~\cite{pdg2020} and $m_{\kzero} = (0.497611 \pm 0.000013)$~\GeVmass~\cite{pdg2020} are the pion and \kzero masses, while $n, \, A, \, B$, and $C$ are fit parameters. For the estimation of the systematic uncertainties second- and third-order polynomial functions were also considered to fit the residual background. The default fitting range used is 0.66--1.10~\GeVmass for every \pt\ interval of each multiplicity class, which was varied for the systematic studies.

\begin{figure}[h!]
\centering
\includegraphics[width=120mm]{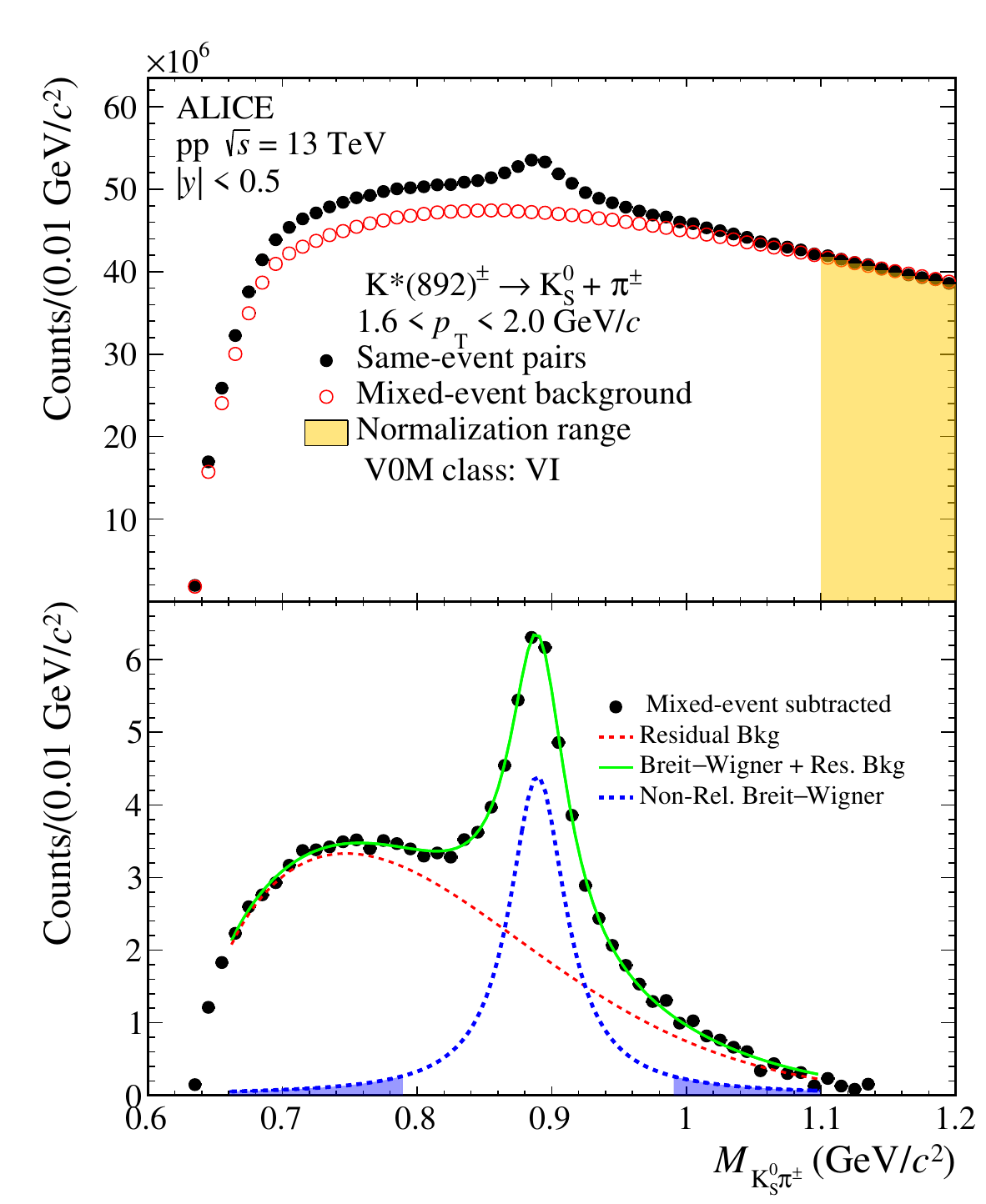}
\caption{The $\kzero\pipm$ invariant mass distribution at $|y|<0.5$ in pp collisions at \thirteen for the $1.6<\pt<2.0$~\GeVc interval in the VI V0M multiplicity class (black full circles), before (upper panel) and after (lower panel) the uncorrelated background subtraction. Statistical uncertainties are shown with error bars. In the upper panel the red open circles describe the background shape evaluated via the event-mixing technique, while the yellow-filled area represents the normalization region (1.1--1.2~\GeVmass) between the mixed-event background and the same-event pair invariant mass distribution. In the lower panel the solid green curve is the result of the fit with Eq.~\ref{eq:eq1}, 
while the dashed red and blue line describe, respectively, the residual background given by Eq.~\ref{eq:eq2} and the non-relativistic Breit--Wigner function. The blue regions represent the tails of the non-relativistic Breit–Wigner fit outside the
integration region.
}
\label{fig:inv_mass}
\end{figure}

The raw yield ($N_{\rm raw}$) of K$^{*\pm}$ is computed by integrating the counts in the measured invariant mass distribution over the peak region ($I_{\rm min} < M_{\kzero\pi} < I_{\rm max}$, where $I_{\rm min} = M_0 - 2\Gamma_0 = 0.79$~\GeVmass and $I_{\rm max} = M_0 + 2\Gamma_0 = 0.99$~\GeVmass), subtracting the integral of the residual background portion estimated by integrating the function $F_{BG}$ over that same interval, and adding the integral of the tails of the non-relativistic Breit--Wigner fit function on both sides of the peak outside the integration region (from $m_{\pipm} + m_{\kzero}$ to $M_0 - 2\Gamma_0$ and from $M_0 + 2\Gamma_0$ to $\infty$), which accounts for about 13$\%$ of the yield. The integration range is another parameter varied for the systematic  uncertainty estimation. Figure~\ref{fig:inv_mass} shows an example of the $\kzero\pipm$ invariant mass distribution before and after the combinatorial background subtraction.

A simulated dataset was also analyzed in order to evaluate the product ($A\times\epsilon_{\rm rec}$) of the detector geometrical acceptance ($A$) and the  K$^{*\pm}$ reconstruction efficiency ($\epsilon_{\rm rec}$). This was generated using particle distributions from the PYTHIA8 Monte Carlo event generator with the Monash 2013 tune~\cite{pythia8} and a GEANT3-based simulation~\cite{geant3} of the ALICE detector. The same event selection, track quality, and V$^0$ decay topology are used for the real and simulated data. The obtained $A\times\epsilon_{\rm rec}$ values as a function of \pt for the different multiplicity classes are consistent with the acceptance$\times$efficiency computed in the inclusive K$^{*\pm}$ analysis and reported in Ref.~\cite{incl_KStarpm} (see Fig.~2 in Ref.~\cite{incl_KStarpm}). 
No dependence on multiplicity is observed, therefore the $A\times\epsilon_{\rm rec}$ computed in the multiplicity-integrated INEL$>0$ class is used as the correction factor for all the multiplicity classes.

For each multiplicity class the differential transverse momentum yield was calculated using Eq.~3 of Ref.~\cite{incl_KStarpm}. The \pt spectra are normalized by the number of accepted events and corrected for the signal-loss ($f_{\rm SL}$) to account for the loss of  K$^{*\pm}$ mesons in non-triggered events. To calculate this correction factor for each multiplicity class, the simulated resonance \pt spectrum before triggering and event selection is divided by the corresponding \pt spectrum after those selections. This correction is a \pt-dependent factor which is peaked at low \pt with the larger deviation from unity in the lowest multiplicity class for $\pt < 1$~\GeVc. In the \pt interval 0--0.4~\GeVc $f_{\rm SL}$ is equal to 1.01 and 1.14 in I and X multiplicity classes, respectively. A correction for the primary vertex reconstruction efficiency ($<1\%$) is also taken into account.

\subsection{Systematic uncertainties}\label{subsec:sys}
The sources of systematic uncertainties considered for the measurement of the K$^{*\pm}$ yields are related to the signal extraction procedure, the track selection and particle identification criteria for the primary pions, the selections on the \kzero decay topology, the primary vertex position range, the imperfect description of the ALICE material budget and of the hadronic interaction cross section in the simulation, and a possible discrepancy between data and Monte Carlo in the TPC--ITS track prolongation efficiency for primary tracks, as outlined in Table~\ref{tab:sys}.
For each source, the average systematic uncertainty over all the multiplicity classes is quoted for three different transverse momentum ranges: low (0--1.2~\GeVc), intermediate (1.2--3.0~\GeVc), and high (3.0--10.0~\GeVc) \pt.
The main contribution arises from the signal extraction procedure (3--4\%) that encloses all the aforementioned variations with respect to the default configuration of the mixed-event normalization range, fit range, residual background function, integration region, and resonance width.

\begin{table}[h!]
  \caption{Sources of systematic uncertainties and the relative multiplicity-averaged values (expressed in \%) on the K$^{*\pm}$ \pt-differential distributions reported for low, intermediate, and high \pt intervals. The contributions from each source are summed in quadrature to obtain the total systematic uncertainty.}
  \label{tab:sys}
\centerline{\scalebox{1}{
\begin{tabular}{|c|ccc|}
\hline 
\pt (\GeVc) &  0 -- 1.2 &	1.2 -- 3.0 & 3.0 -- 10.0\\
\hline
Signal extraction (\%)         &	 3.9  & 	2.8  &	3.2 \\
Primary pion selection (\%)	   &   1.5  & 	0.9  &	1.4 \\
\kzero selection (\%)     	   &   1.8  & 	1.3  &	1.7 \\
Primary vertex (\%)	           &   1.3  & 	1.3  &	1.7 \\
Material budget (\%)           &	 3.0  & 	1.5  &	0.7 \\
Hadronic interaction (\%) 	   &   1.1  & 	1.1  &	0.5 \\
Global tracking efficiency (\%) &	 1.3  & 	2.1  &	2.5 \\
\hline
Total (\%)                     &   5.9  &   4.5  &  5.0 \\
\hline
\end{tabular}
}}
\end{table}

The systematic uncertainties related to the primary pion identification, to the reconstruction of secondary tracks and \kzero decay vertices, and to the selected range of primary vertex position are estimated by varying the corresponding selection criteria both in the data and in Monte Carlo simulations. The uncertainty due to the primary pion selection is about 1--1.5\%, estimated by repeating the analysis varying the track-quality selection criteria. To study the effect of PID, the selection criteria based on the TPC energy loss were varied with respect to the default setting previously described. Specifically, PID selection criteria of 2.5$\sigma_{\rm TPC}$~and 4$\sigma_{\rm TPC}$~were used.

The contribution to the systematic uncertainty originating from the \kzero\ reconstruction ranges from 1.3\% to 1.8\%. It was estimated by varying the selection criteria on the track quality and PID for the secondary pions, the DCA between the \kzero decay products and to the primary vertex, the cosine of pointing angle, and the V$^{0}$ decay radius, invariant mass, rapidity, and proper lifetime. As in the case of the inclusive K$^{*\pm}$ analysis~\cite{incl_KStarpm}, the total systematic uncertainty due to the \kzero\ reconstruction and selection is lower than the corresponding uncertainty on the K$^{\pm}$ measurement for the same collision system and energy~\cite{charged_hadr}. In particular, the large contribution to the uncertainty due to primary track and PID selection for K$^{\pm}$ is here avoided thanks to the \kzero topology.
The dependence of the results on the position of the primary vertex along the beam axis was tested by varying the vertex selection window by $\pm 2$~cm with respect to the default value (i.e. 10~cm). A systematic uncertainty of 1.3--1.7\% was estimated on the measured K${^{*\pm}}$ yields.
Additional uncertainties are those related to the knowledge of the ALICE material budget (varying from 3\% for \pt $<$1.2~\GeVc to 1.5\% for 1.2$<$\pt $<3$~\GeVc interval and 0.7\% at higher \pt) and of the hadronic interaction cross section ($1.1\%$ for \pt $<3$~\GeVc and 0.5\% at higher \pt). 
These effects are evaluated by combining the uncertainties for a pion and a \kzero, determined as in Refs.~\cite{res_lam, multi-strange}, according to the kinematics of the decay. An additional uncertainty originates from an imperfect description in the simulation of the probability of prolonging a TPC track to the ITS. This systematic uncertainty was estimated by comparing the matching efficiency for primary particles in data and simulations and ranges from 1.3\% to 2.5\%.

The total K$^{*\pm}$ systematic uncertainty for the considered \pt intervals is obtained by adding in quadrature the contributions from the different sources. Its value is about 4.5--6\%, while the total systematic uncertainty obtained for the K$^{*0}$ measurements at specific low (0.2~\GeVc), intermediate (2.2~\GeVc), and high (6.5~\GeVc) \pt ranges from 8\% to 12\%. 
The systematic uncertainties were studied independently for INEL$>$0 events and in each individual multiplicity class, to separate the sources that are correlated and uncorrelated across multiplicity intervals, following the same procedure as in Refs.~\cite{multi-strange,KStar_pp7}. In particular, signal extraction, primary pion selection, \kzero reconstruction and selection, and primary vertex are uncorrelated sources, whereas TPC--ITS matching efficiency, material budget, and hadronic cross sections are correlated among different event multiplicity classes.

\section{Results}\label{sec:results}
\begin{figure}[h!]  \centering
    \includegraphics[scale=0.5]{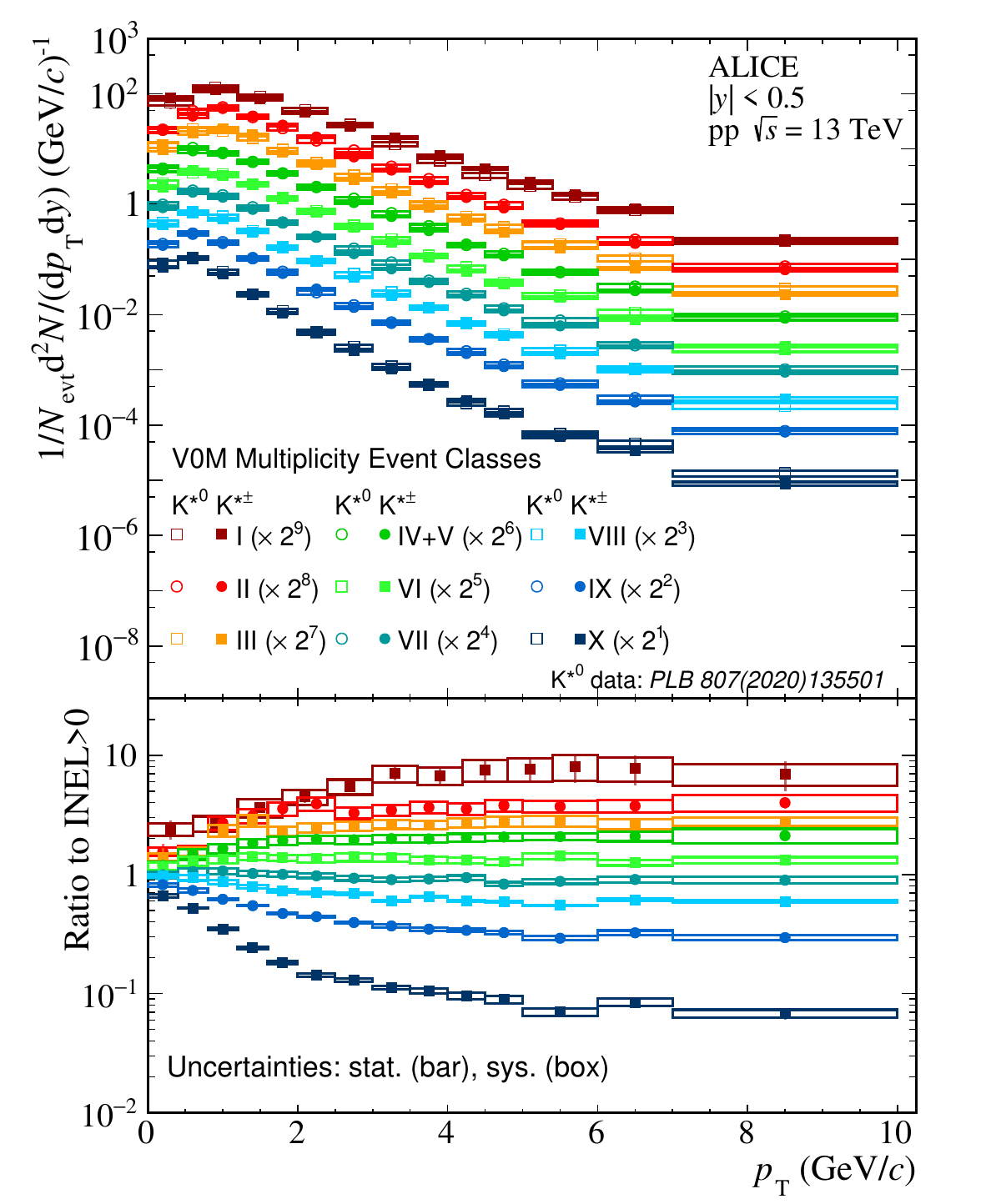}
    \caption{The K$^{*\pm}$ and K$^{*0}$~\cite{KStar_pp13} \pt distributions in pp collisions at \thirteen for the different multiplicity classes studied, scaled by the indicated factors. Lower panel: ratios of the \pt spectra in each multiplicity class to the multiplicity-integrated INEL$>0$ spectrum.
}
    
    \label{fig:diff_spectra}
\end{figure}

The \pt-differential yields of K$^{*\pm}$ in the various multiplicity classes for pp collisions at \thirteen, as well as their ratios to the inclusive INEL$>$0 distributions, are shown in Fig.~\ref{fig:diff_spectra}. In this figure, the K$^{*0}$ distributions measured for the same system and energy~\cite{KStar_pp13} are also reported for comparison. 

The yields and \pt distributions of the charged and neutral mesons are consistent within the uncertainties. As for the K$^{*0}$, the \pt-differential distributions of the K$^{*\pm}$ get harder with increasing multiplicity and for higher multiplicity classes the \pt value at which the yield is maximum shifts towards higher momenta. This trend is qualitatively similar to the behavior observed in heavy-ion collisions~\cite{rho, KStar5, KStarPbPb, KStar2.1, sigma, lambda} and attributed to a collective radial expansion of the system, although as mentioned in Sec.~\ref{sec:intro}, the color reconnection mechanism seems to mimic collective-like effects in small systems~\cite{CR}. The lower panel of Fig.~\ref{fig:diff_spectra} shows the ratios of the \pt distribution measured in each individual multiplicity class to the inclusive spectrum (INEL$>0$ class). The hardening of the spectra is significant for $\pt \lesssim 4$~\GeVc, while at higher \pt the ratios are flat for all multiplicity classes.
The results have many qualitative similarities to those obtained for longer lived hadrons in pp collisions at \thirteen~\cite{multi-strange, charged_hadr}, and are consistent with previous measurements of hadronic resonances in small collisions systems~\cite{KStar_pPb, KStar_pp13,pPbSigandXi, Lambda_pp_pPb}.

The \pt-integrated K$^{*\pm}$ yields per-event (corresponding to 1/$N_{\rm INEL} \times$ d$N$/d$y$, hereby denoted as d$N$/d$y$ for brevity) and the mean transverse momenta \meanpt in the different multiplicity classes are determined by integrating and averaging the transverse momentum spectra over the measured range ($0 < \pt < 10$~\GeVc).

\begin{figure}[h!]
\centering
\includegraphics[width=120mm]{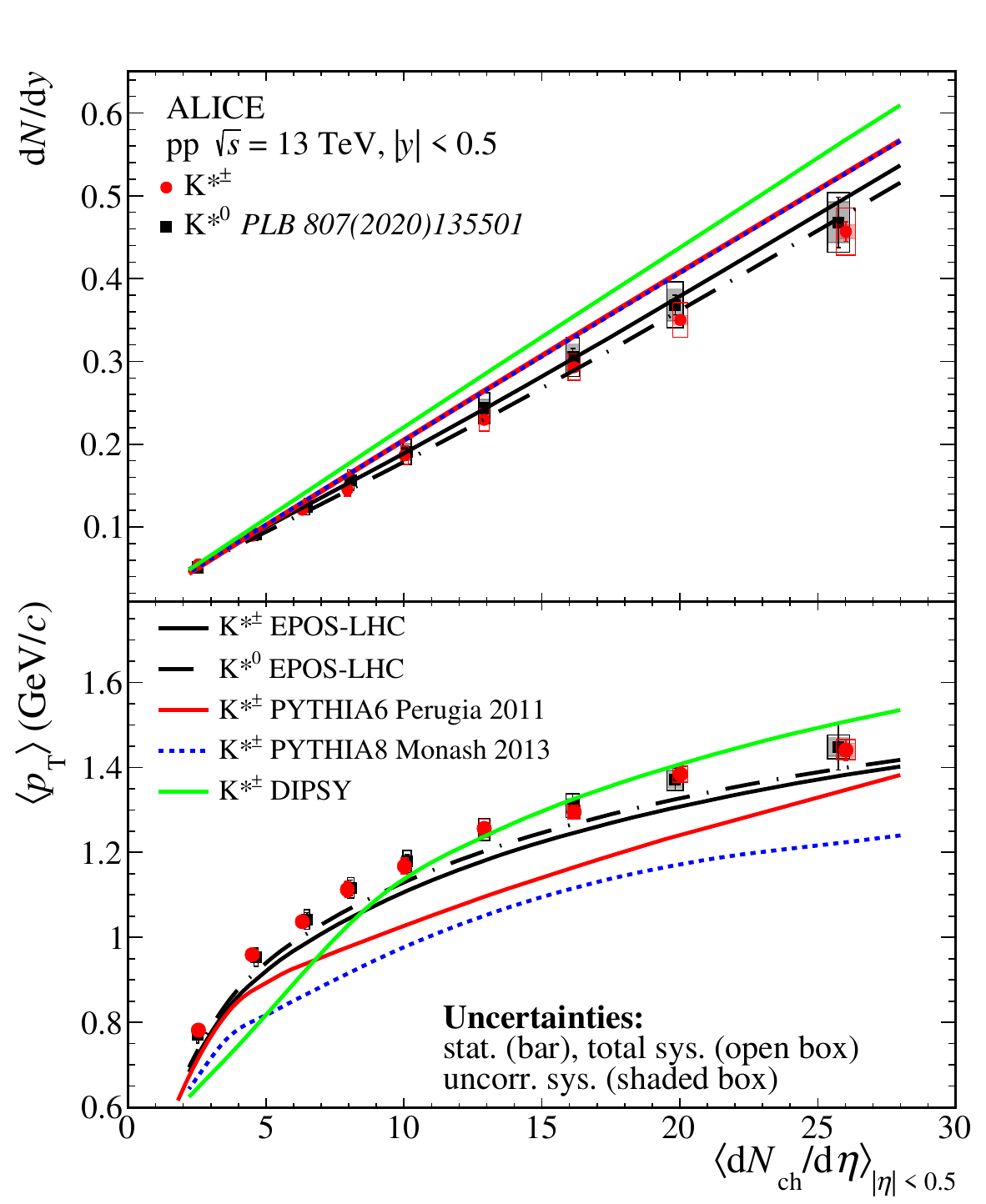}
\caption{The per-event \pt-integrated yields d$N$/d$y$ (upper panel) and mean transverse momenta \meanpt (lower panel) for K$^{*\pm}$ and K$^{*0}$~\cite{KStar_pp13} as a function of the average charged-particle multiplicity density \avdndetamid. Bars represent statistical uncertainties, open boxes represent total systematic uncertainties, and shaded boxes are the systematic uncertainties uncorrelated with multiplicity.  Symbol dimension represents the average charged-particle density uncertainty.  The measurements are also compared to predictions from different event generators~\cite{pythia6, pythia8, epos-lhc, dipsy}. 
}
\label{fig:dndy_and_meanpt}
\end{figure}

Figure~\ref{fig:dndy_and_meanpt} shows the \pt-integrated yields and the mean \pt of the K$^{*\pm}$ and K$^{*0}$ as a function of the average charged-particle multiplicity density \avdndetamid, together with predictions from different event generators (PYTHIA6, Perugia 2011 tune~\cite{pythia6}; PYTHIA8, Monash 2013 tune~\cite{pythia8}; EPOS-LHC~\cite{epos-lhc}; and DIPSY~\cite{dipsy}).
The \avdndetamid values for the multiplicity classes utilized for charged and neutral K$^{*}$ refer to different datasets and are consistent within the estimated uncertainties.  
The d$N$/d$y$ and \meanpt of charged and neutral K$^*$ are in agreement within the systematic and statistical uncertainties. 

PYTHIA is a general purpose Monte Carlo event generator used for the description of several types of high energy collisions. The Lund string model is used in PYTHIA to convert the partons into hadrons~\cite{lund}. The QCD force between partons is described in terms of phenomenological strings which break to produce hadrons. The PYTHIA event generator has been updated over the years implementing different new tunings. In particular, in PYTHIA tunes including the color reconnection effect~\cite{pythia8CR} the string connecting two partons follows the evolution of the partonic endpoints resulting in a common boost of the string fragments (i.e. the hadrons), mimicking the collective flow that affects the hadron spectra in heavy-ion collisions~\cite{CR}. Furthermore, color reconnection suppresses the relative production of meson resonances such as  K$^*$~\cite{pythia8CR} reducing the K$^*/$K ratio.

EPOS-LHC is a Monte Carlo event generator based on the parton-based Gribov--Regge theory~\cite{gribov}. It applies a common approach for pp, p--A, and A--A collisions using the same formalism but different parameterizations. This model is based on the core--corona separation, where the string segments are grouped into two different regions: a high density area known as \lq\lq core" surrounded by a low density one, referred as \lq\lq corona". 
The core represents a region with a high density of string segments that is larger than some critical density for which the hadronization is treated collectively and the corona is the region with a lower density of string segments for which the hadronization is treated non collectively.
The modeling of the collective flow of the core depends on the collision system. It varies from a quick expansion of a very dense system compressed in a small volume, as could happen in high multiplicity pp collisions, to a large volume medium as the one created in heavy-ion collisions. 
The high density core can be formed by the overlap of the string fragments as a consequence of multiple parton interactions in the case of pp collisions or for multiple nucleon interactions for A--A collisions. Furthermore, EPOS-LHC doesn’t include any hadronic afterburner which could take care to reproduce rescattering and regeneration effects in the hadronic phase.

DIPSY~\cite{dipsy0,dipsy,dipsy1} is a Monte-Carlo event generator  based on Mueller's dipole cascade model which is a formulation of leading logaritimic BFKL (Baltisky-Fadin–Kuraev–Lipatov) evolution equation~\cite{Kuraev:1977fs,Balitsky:1978ic,Fadin:1975} in transverse coordinate space. It is used to simulate complete minimum-bias non-diffractive hadronic collision events and to investigate  soft and semi-hard processes involved in both hadronic and heavy-ion collisions.

The measured d$N$/d$y$ values exhibit an approximately linear increase with increasing \avdndetamid well described by EPOS-LHC, while PYTHIA and DIPSY calculations tend to overestimate the data. The observed differences between charged and neutral K$^*$ production in EPOS-LHC are due to an overestimation of the isospin symmetry breaking in the string fragmentation~\cite{private_communication}. The $\langle$\pt$\rangle$ increases with the event multiplicity too, showing a sign of saturation in high-multiplicity events. The PYTHIA6 - Perugia2011  and PYTHIA8 - Monash 2013  predictions have a consistent trend similar to the data although they largely underpredict them. 
The EPOS-LHC predictions are consistent with the observed trend although the values are slightly underestimated. DIPSY, on the other hand, shows a more pronounced increase in mean \pt from low to high multiplicity than data. 

As already discussed in Sec.~\ref{sec:intro} the ratio of \pt-integrated resonance yields to their ground states is an important tool to verify the presence of modifications in resonance production and its dependence on the system size. In heavy-ion collisions a suppression of the ratio of the yield of resonances relative to ground-state particles is interpreted as the dominance of the rescattering mechanism over regeneration processes. For similarity, in pp collisions the observation of a suppression could be a hint of the presence of a hadronic phase also in elementary collisions. 

\begin{figure}[h!]
\centering
\includegraphics[scale=0.8]{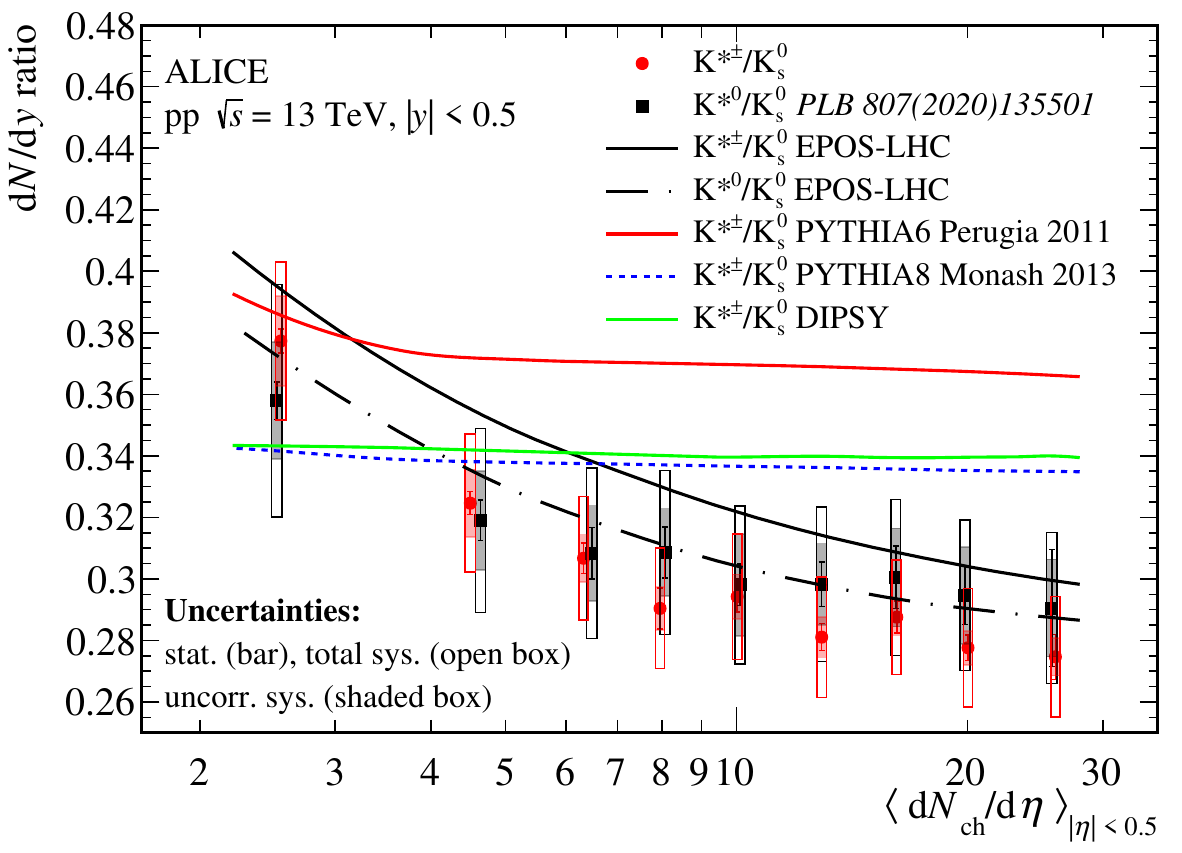}
\caption{Ratios of K$^{*\pm}$/\kzero and K$^{*0}$/\kzero~\cite{KStar_pp13} yields in pp collisions at \thirteen as a function of the average charged-particle multiplicity density at  midrapidity \avdndetamid. Bars represent statistical uncertainties, open boxes represent total systematic uncertainties, and shaded boxes are the systematic uncertainties uncorrelated with multiplicity. The measurements are also compared to predictions from the  PYTHIA6, PYTHIA8, EPOS-LHC, and DIPSY event generators~\cite{pythia6,pythia8,epos-lhc,dipsy}. The prediction from EPOS-LHC for K$^{*0}$/\kzero ratio distribution reported in~\cite{KStar_pp13} is also shown.}

\label{fig:ratio}
\end{figure}
Figure~\ref{fig:ratio} shows the multiplicity dependence of the K$^{*\pm}$/\kzero ratio in pp collisions at \thirteen, compared to the K$^{*0}$/\kzero one, with the \kzero data from Ref.~\cite{multi-strange}. The decreasing trend passing from low to high multiplicity pp collisions, already outlined by the K$^{*0}$ analysis~\cite{KStar_pp13}, is confirmed by the K$^{*\pm}$ results. The K$^{*\pm}$/\kzero\, results provide a more significant evidence for this suppression thanks to the smaller uncertainties as compared to the previous K$^{*0}$ results. Indeed, the suppression of the K$^{*0}$/\kzero ratio between the highest and lowest multiplicity classes was measured at a significance level of about $2\sigma$~\cite{KStar_pp13}. Due to increased precision, the K$^{*\pm}$/\kzero suppression between the highest and lowest multiplicity classes is measured at a significance level of $7\sigma$, taking into account the multiplicity-uncorrelated uncertainties. This result represents the first observation of a clear K$^*$/K suppression in a small collision system. 
The yield ratios shown in Fig.~\ref{fig:ratio} are also compared to the same models considered for the d$N$/d$y$ and the \meanpt. 
The prediction of EPOS-LHC for  K$^{*0}$/\kzero reported in Ref.~\cite{KStar_pp13} is also shown.
EPOS-LHC provides good agreement with the measured data, well reproducing the decreasing trend. The  K$^{*0}$/\kzero distribution is better reproduced. The PYTHIA6, PYTHIA8, and DIPSY models tend to overestimate the ratios at high multiplicities and exhibit a fairly flat trend. It is worth to note that none of these event generators considers the evolution of a hadronic phase and of eventually hadronic interactions of generated particles. Moreover, EPOS-LHC without the use of any hadronic afterburner is able to reproduce fairly well the measured decreasing trend of the K*/K ratio with the multiplicity. This is a consequence of the varying relative contributions of core and corona importance across different multiplicity classes, as well as the distinct hadronization scenarios associated with each region. Core production becomes dominant in high-multiplicity events, whereas corona production plays a more significant role at low multiplicities. In the core, hadronization proceeds collectively due to the high density of string segments, while it follows a non-collective behavior in the corona region.

\begin{figure}[h!]
\centering
{\includegraphics[scale=0.8]{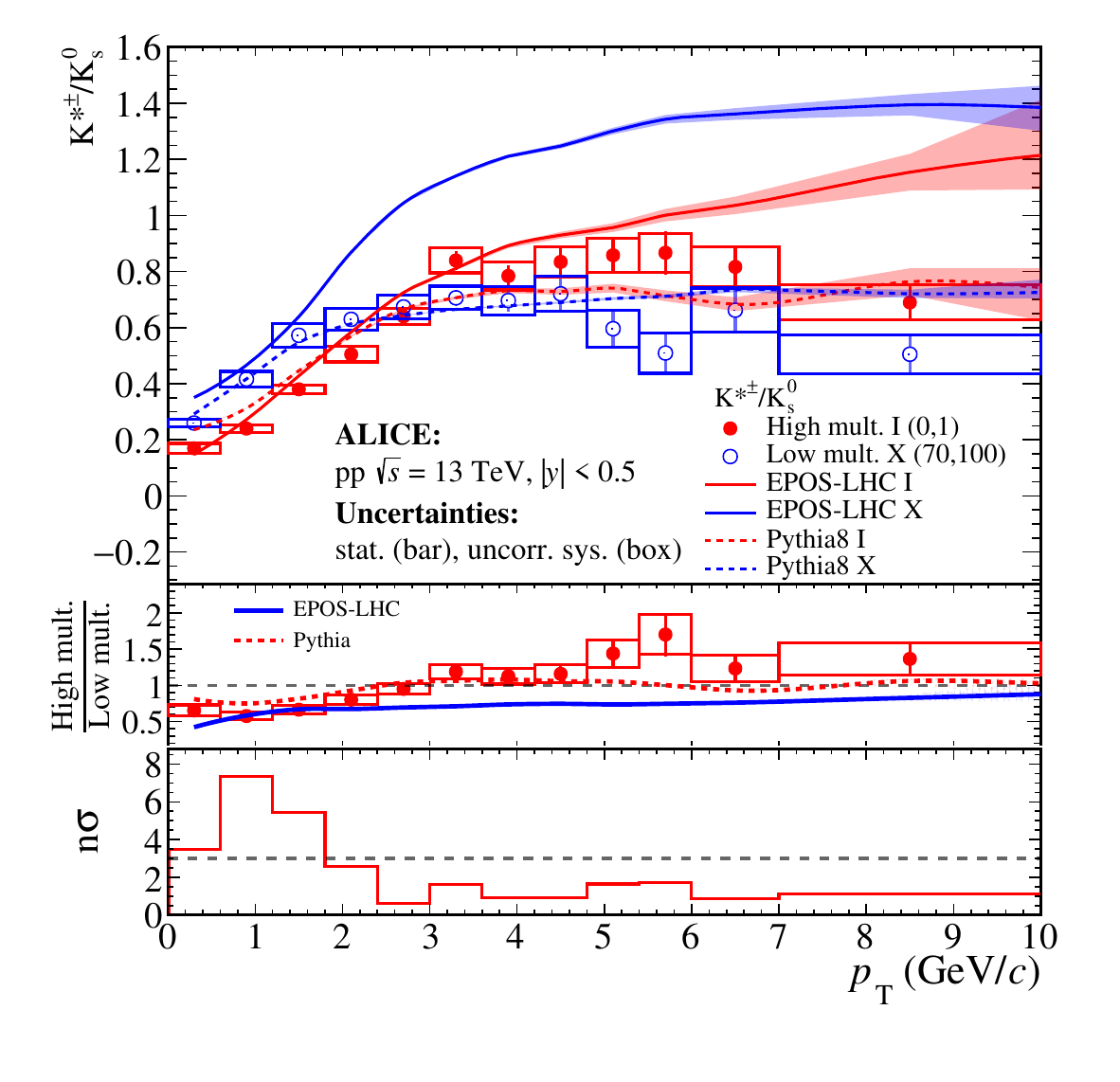}}
\caption{Upper panel: ratios of K$^{*\pm}$/\kzero as a function of \pt for low (X) and high (I) multiplicity classes. Predictions for PYTHIA8 - Monash2013 (dashed lines)  and EPOS-LHC (continuous lines) are also reported.
Middle panel: the highest multiplicity K$^{*\pm}$/\kzero\ ratio divided by the  lowest multiplicity one (double ratio). Lower panel: significance of the deviation of the double ratio from unity. The dashed black line indicates a deviation at the $3\sigma$~level. Bars represent the statistical uncertainties, while boxes represent the part of the systematic uncertainty that is uncorrelated between the multiplicity classes.}
\label{fig:double_ratio}
\end{figure}

The \pt dependence of the K$^{*\pm}$/\kzero ratio is shown in Fig.~\ref{fig:double_ratio} for low and high multiplicity classes (X and I, respectively). The ratios increase with increasing \pt at low transverse momentum and become flat for $\pt \gtrsim 3$~\GeVc, with the K$^{*\pm}$/\kzero ratio computed in the highest multiplicity class (I) being consistent with the one measured in the lowest  multiplicity class (X). This behavior is consistent with the results obtained for K$^{*0}$ in pp~\cite{KStar_pp13} and Pb--Pb~\cite{KStar2.1} collisions.
The prediction of the ratios K$^{*\pm}$/\kzero as a function of \pt for the two multiplicity classes is also shown.  PYTHIA8 - Monash 2013  well reproduces the ratio obtained in low multiplicity events, but for high multiplicity events overestimates the ratio for \pt lower that 3~\GeVc while underestimates the same ratio for \pt larger than 3~\GeVc. EPOS-LHC well reproduces the yields measured in high multiplicity events for \pt less than 4 \GeVc, while overestimates the data in low multiplicity events especially for \pt larger than 1.5~\GeVc.  To quantify the multiplicity dependence of the K$^{*\pm}$/\kzero ratio in pp collisions at \thirteen, the middle panel of Fig.~\ref{fig:double_ratio} shows the double ratio, i.e. the high-multiplicity K$^{*\pm}$/\kzero \pt-dependent ratio divided by that in the lowest multiplicity class. This is close to unity for $\pt \gtrsim 2.5$~\GeVc while for $\pt \lesssim 2.5$~\GeVc a suppression from low to high-multiplicity collisions is appreciable. This is quantified in the lower panel where the significance of the deviations of the double ratio from unity is shown. For $\pt \lesssim 2$~\GeVc the deviation is measured at more than $3\sigma$~level. In heavy-ion collisions a suppression at low \pt is commonly understood as an indication of the presence of a hadronic phase and the dominance of rescattering processes over the regeneration one. For similarity, in pp collisions, the observed suppression at low \pt could be interpreted as a hint of  a hadronic phase with a short lifetime~\cite{lifetime_RHIC}. 
However, QCD event generators are able to reproduce the main characteristics of the K$^{*\pm}$ production without including a hadronic phase. EPOS-LHC without any hadronic afterburner is able to reproduce fairly well the measured decreasing trend of the K*/K ratios versus multiplicity, while PYTHIA8- Monash 2013 tune is able to reproduce the shape of the \pt distribution of the ratio.

\section{Conclusion}\label{sec:conclusion}
Precise measurements of the K$^{*\pm}$ resonance production in pp collisions at \thirteen in different multiplicity classes have been reported.
The results have many qualitative similarities to those reported for longer-lived hadrons in the same collision system~\cite{KStar_pp7, sys_Ks, multi-strange, charged_hadr} and are consistent with previous results of K$^{*0}$ at the same collision energy~\cite{KStar_pp13}.
For $\pt \lesssim 4$~\GeVc the K$^{*\pm}$ differential \pt spectra get harder with increasing multiplicity, while for higher \pt values the same spectral shape can be noticed for all multiplicity classes. The multiplicity-dependent behavior in the low \pt region is qualitatively similar to the trend observed in heavy-ion collisions~\cite{rho, KStar5, KStarPbPb, KStar2.1, sigma, lambda}, where this effect is attributed to the collective expansion of the fireball. However, for small collision systems, in PYTHIA event generator color reconnection mechanism can mimic a flow-like pattern~\cite{CR}.

Both the \pt-integrated yields and the mean transverse momenta are computed from the measured \pt spectra as a function of the average charged-particle multiplicity density and exhibit an increase (approximately linear for integrated yields) from low to high multiplicity. The multiplicity dependence of the yields  is fairly well described by EPOS-LHC, which underpredicts the 
$\langle$\pt$\rangle$ values, reproducing the general increasing trend with the event multiplicity. 

Considering the \pt-integrated particle yield ratios, a suppression of the K$^{*\pm}$/\kzero ratio at a $\sim 7\sigma$~level is observed passing from low to high multiplicity pp collisions. This is the first observation of K$^*$/K suppression in small collision systems. Previously, a hint of suppression at 2$\sigma$~level was observed in  K$^{*0}$/\kzero~ratio in \pp~collisions at \s~=~13~TeV~\cite{KStar_pp13}. The \pt-differential K$^{*\pm}$/\kzero  ratio at high multiplicity to the low-multiplicity one helps to quantify the observed decrease of the particle ratios. For $\pt \lesssim 2$~\GeVc, the K$^{*\pm}$/\kzero double ratio deviates from unity by more than $3\sigma$~suggesting that the observed suppression occurs mainly at low \pt. The K$^{*\pm}$/\kzero multiplicity evolution is consistent with the hypothesis of significance of rescattering process of the short-lived resonances decay particles which is expected to have greater strength for low-\pt resonances. This trend, for similarity with heavy-ion collisions, could suggest the presence of a short hadronic phase in small systems. However, considering that the distributions of the K*/K ratios as a multiplicity and \pt function are rather well reproduced by QCD generators without any hadronic phase, the presence of a hadronic phase in pp collisions is still an open question. 


\newenvironment{acknowledgement}{\relax}{\relax}
\begin{acknowledgement}
\section*{Acknowledgements}
\input{fa_2025-05-20_Opt_C.tex}
\end{acknowledgement}

\bibliographystyle{utphys}   
\bibliography{bibliography}

\newpage
\appendix

%
%

\section{The ALICE Collaboration}
\label{app:collab}
\input{Alice_Authorlist_2025-05-20_Opt_C.tex}

\end{document}

%% file: commands.tex
%

\newcommand{\pp}           {pp\xspace}
\newcommand{\ppbar}        {\mbox{$\mathrm {p\overline{p}}$}\xspace}
\newcommand{\XeXe}         {\mbox{Xe--Xe}\xspace}
\newcommand{\PbPb}         {\mbox{Pb--Pb}\xspace}
\newcommand{\pA}           {\mbox{pA}\xspace}
\newcommand{\pPb}          {\mbox{p--Pb}\xspace}
\newcommand{\AuAu}         {\mbox{Au--Au}\xspace}
\newcommand{\dAu}          {\mbox{d--Au}\xspace}

\newcommand{\s}            {\ensuremath{\sqrt{s}}\xspace}
\newcommand{\snn}          {\ensuremath{\sqrt{s_{\mathrm{NN}}}}\xspace}
\newcommand{\pt}           {\ensuremath{p_{\rm T}}\xspace}
\newcommand{\meanpt}       {$\langle p_{\mathrm{T}}\rangle$\xspace}
\newcommand{\ycms}         {\ensuremath{y_{\rm CMS}}\xspace}
\newcommand{\ylab}         {\ensuremath{y_{\rm lab}}\xspace}
\newcommand{\etarange}[1]  {\mbox{$\left | \eta \right |~<~#1$}}
\newcommand{\yrange}[1]    {\mbox{$\left | y \right |~<~#1$}}
\newcommand{\dndy}         {\ensuremath{\mathrm{d}N_\mathrm{ch}/\mathrm{d}y}\xspace}
\newcommand{\dndeta}       {\ensuremath{\mathrm{d}N_\mathrm{ch}/\mathrm{d}\eta}\xspace}
\newcommand{\avdndeta}     {\ensuremath{\langle\dndeta\rangle}\xspace}
\newcommand{\avdndetamid}     {\ensuremath{\langle\dndeta\rangle}$_{|\eta|<0.5}$\xspace}
\newcommand{\dNdy}         {\ensuremath{\mathrm{d}N_\mathrm{ch}/\mathrm{d}y}\xspace}
\newcommand{\Npart}        {\ensuremath{N_\mathrm{part}}\xspace}
\newcommand{\Ncoll}        {\ensuremath{N_\mathrm{coll}}\xspace}
\newcommand{\dEdx}         {\ensuremath{\textrm{d}E/\textrm{d}x}\xspace}
\newcommand{\RpPb}         {\ensuremath{R_{\rm pPb}}\xspace}

\newcommand{\nineH}        {$\sqrt{s}~=~0.9$~Te\kern-.1emV\xspace}
\newcommand{\seven}        {$\sqrt{s}~=~7$~Te\kern-.1emV\xspace}
\newcommand{\twoH}         {$\sqrt{s}~=~0.2$~Te\kern-.1emV\xspace}
\newcommand{\twosevensix}  {$\sqrt{s}~=~2.76$~Te\kern-.1emV\xspace}
\newcommand{\five}         {$\sqrt{s}~=~5.02$~Te\kern-.1emV\xspace}
\newcommand{\thirteen}     {$\sqrt{s}~=~13$~Te\kern-.1emV\xspace}
\newcommand{\twosevensixnn}{$\sqrt{s_{\mathrm{NN}}}~=~2.76$~Te\kern-.1emV\xspace}
\newcommand{\fivenn}       {$\sqrt{s_{\mathrm{NN}}}~=~5.02$~Te\kern-.1emV\xspace}
\newcommand{\LT}           {L{\'e}vy-Tsallis\xspace}
\newcommand{\GeVc}         {Ge\kern-.1emV/$c$\xspace}
\newcommand{\MeVc}         {Me\kern-.1emV/$c$\xspace}
\newcommand{\TeV}          {Te\kern-.1emV\xspace}
\newcommand{\GeV}          {Ge\kern-.1emV\xspace}
\newcommand{\MeV}          {Me\kern-.1emV\xspace}
\newcommand{\GeVmass}      {Ge\kern-.1emV/$c^2$\xspace}
\newcommand{\MeVmass}      {Me\kern-.1emV/$c^2$\xspace}
\newcommand{\lumi}         {\ensuremath{\mathcal{L}}\xspace}

\newcommand{\ITS}          {\rm{ITS}\xspace}
\newcommand{\TOF}          {\rm{TOF}\xspace}
\newcommand{\ZDC}          {\rm{ZDC}\xspace}
\newcommand{\ZDCs}         {\rm{ZDCs}\xspace}
\newcommand{\ZNA}          {\rm{ZNA}\xspace}
\newcommand{\ZNC}          {\rm{ZNC}\xspace}
\newcommand{\SPD}          {\rm{SPD}\xspace}
\newcommand{\SDD}          {\rm{SDD}\xspace}
\newcommand{\SSD}          {\rm{SSD}\xspace}
\newcommand{\TPC}          {\rm{TPC}\xspace}
\newcommand{\TRD}          {\rm{TRD}\xspace}
\newcommand{\VZERO}        {\rm{V0}\xspace}
\newcommand{\VZEROA}       {\rm{V0A}\xspace}
\newcommand{\VZEROC}       {\rm{V0C}\xspace}
\newcommand{\Vdecay} 	   {\ensuremath{V^{0}}\xspace}

\newcommand{\ee}           {\ensuremath{e^{+}e^{-}}} 
\newcommand{\pip}          {\ensuremath{\pi^{+}}\xspace}
\newcommand{\pim}          {\ensuremath{\pi^{-}}\xspace}
\newcommand{\pipm}          {\ensuremath{\pi^{\pm}}\xspace}
\newcommand{\kap}          {\ensuremath{\rm{K}^{+}}\xspace}
\newcommand{\kam}          {\ensuremath{\rm{K}^{-}}\xspace}
\newcommand{\pbar}         {\ensuremath{\rm\overline{p}}\xspace}
\newcommand{\kzero}        {\ensuremath{{\rm K}^{0}_{\rm{S}}}\xspace}
\newcommand{\lmb}          {\ensuremath{\Lambda}\xspace}
\newcommand{\almb}         {\ensuremath{\overline{\Lambda}}\xspace}
\newcommand{\Om}           {\ensuremath{\Omega^-}\xspace}
\newcommand{\Mo}           {\ensuremath{\overline{\Omega}^+}\xspace}
\newcommand{\X}            {\ensuremath{\Xi^-}\xspace}
\newcommand{\Ix}           {\ensuremath{\overline{\Xi}^+}\xspace}
\newcommand{\Xis}          {\ensuremath{\Xi^{\pm}}\xspace}
\newcommand{\Oms}          {\ensuremath{\Omega^{\pm}}\xspace}
\newcommand{\degree}       {\ensuremath{^{\rm o}}\xspace}

%% file: fa_2025-05-20_Opt_C.tex

The ALICE Collaboration would like to thank all its engineers and technicians for their invaluable contributions to the construction of the experiment and the CERN accelerator teams for the outstanding performance of the LHC complex.
The ALICE Collaboration gratefully acknowledges the resources and support provided by all Grid centres and the Worldwide LHC Computing Grid (WLCG) collaboration.
The ALICE Collaboration acknowledges the following funding agencies for their support in building and running the ALICE detector:
A. I. Alikhanyan National Science Laboratory (Yerevan Physics Institute) Foundation (ANSL), State Committee of Science and World Federation of Scientists (WFS), Armenia;
Austrian Academy of Sciences, Austrian Science Fund (FWF): [M 2467-N36] and Nationalstiftung f\"{u}r Forschung, Technologie und Entwicklung, Austria;
Ministry of Communications and High Technologies, National Nuclear Research Center, Azerbaijan;
Rede Nacional de Física de Altas Energias (Renafae), Financiadora de Estudos e Projetos (Finep), Funda\c{c}\~{a}o de Amparo \`{a} Pesquisa do Estado de S\~{a}o Paulo (FAPESP) and The Sao Paulo Research Foundation  (FAPESP), Brazil;
Bulgarian Ministry of Education and Science, within the National Roadmap for Research Infrastructures 2020-2027 (object CERN), Bulgaria;
Ministry of Education of China (MOEC) , Ministry of Science \& Technology of China (MSTC) and National Natural Science Foundation of China (NSFC), China;
Ministry of Science and Education and Croatian Science Foundation, Croatia;
Centro de Aplicaciones Tecnol\'{o}gicas y Desarrollo Nuclear (CEADEN), Cubaenerg\'{\i}a, Cuba;
Ministry of Education, Youth and Sports of the Czech Republic, Czech Republic;
The Danish Council for Independent Research | Natural Sciences, the VILLUM FONDEN and Danish National Research Foundation (DNRF), Denmark;
Helsinki Institute of Physics (HIP), Finland;
Commissariat \`{a} l'Energie Atomique (CEA) and Institut National de Physique Nucl\'{e}aire et de Physique des Particules (IN2P3) and Centre National de la Recherche Scientifique (CNRS), France;
Bundesministerium f\"{u}r Bildung und Forschung (BMBF) and GSI Helmholtzzentrum f\"{u}r Schwerionenforschung GmbH, Germany;
General Secretariat for Research and Technology, Ministry of Education, Research and Religions, Greece;
National Research, Development and Innovation Office, Hungary;
Department of Atomic Energy Government of India (DAE), Department of Science and Technology, Government of India (DST), University Grants Commission, Government of India (UGC) and Council of Scientific and Industrial Research (CSIR), India;
National Research and Innovation Agency - BRIN, Indonesia;
Istituto Nazionale di Fisica Nucleare (INFN), Italy;
Japanese Ministry of Education, Culture, Sports, Science and Technology (MEXT) and Japan Society for the Promotion of Science (JSPS) KAKENHI, Japan;
Consejo Nacional de Ciencia (CONACYT) y Tecnolog\'{i}a, through Fondo de Cooperaci\'{o}n Internacional en Ciencia y Tecnolog\'{i}a (FONCICYT) and Direcci\'{o}n General de Asuntos del Personal Academico (DGAPA), Mexico;
Nederlandse Organisatie voor Wetenschappelijk Onderzoek (NWO), Netherlands;
The Research Council of Norway, Norway;
Pontificia Universidad Cat\'{o}lica del Per\'{u}, Peru;
Ministry of Science and Higher Education, National Science Centre and WUT ID-UB, Poland;
National Research Foundation of Korea (NRF), Republic of Korea;
Ministry of Education and Scientific Research, Institute of Atomic Physics, Ministry of Research and Innovation and Institute of Atomic Physics and Universitatea Nationala de Stiinta si Tehnologie Politehnica Bucuresti, Romania;
Ministerstvo skolstva, vyskumu, vyvoja a mladeze SR, Slovakia;
National Research Foundation of South Africa, South Africa;
Swedish Research Council (VR) and Knut \& Alice Wallenberg Foundation (KAW), Sweden;
European Organization for Nuclear Research, Switzerland;
Suranaree University of Technology (SUT), National Science and Technology Development Agency (NSTDA) and National Science, Research and Innovation Fund (NSRF via PMU-B B05F650021), Thailand;
Turkish Energy, Nuclear and Mineral Research Agency (TENMAK), Turkey;
National Academy of  Sciences of Ukraine, Ukraine;
Science and Technology Facilities Council (STFC), United Kingdom;
National Science Foundation of the United States of America (NSF) and United States Department of Energy, Office of Nuclear Physics (DOE NP), United States of America.
In addition, individual groups or members have received support from:
Czech Science Foundation (grant no. 23-07499S), Czech Republic;
FORTE project, reg.\ no.\ CZ.02.01.01/00/22\_008/0004632, Czech Republic, co-funded by the European Union, Czech Republic;
European Research Council (grant no. 950692), European Union;
Deutsche Forschungs Gemeinschaft (DFG, German Research Foundation) ``Neutrinos and Dark Matter in Astro- and Particle Physics'' (grant no. SFB 1258), Germany;
ICSC - National Research Center for High Performance Computing, Big Data and Quantum Computing and FAIR - Future Artificial Intelligence Research, funded by the NextGenerationEU program (Italy).

%% file: Alice_Authorlist_2025-05-20_Opt_C.tex
\begin{flushleft} 
\small

I.J.~Abualrob\,\orcidlink{0009-0005-3519-5631}\,$^{\rm 113}$, 
S.~Acharya\,\orcidlink{0000-0002-9213-5329}\,$^{\rm 50}$, 
G.~Aglieri Rinella\,\orcidlink{0000-0002-9611-3696}\,$^{\rm 32}$, 
L.~Aglietta\,\orcidlink{0009-0003-0763-6802}\,$^{\rm 24}$, 
M.~Agnello\,\orcidlink{0000-0002-0760-5075}\,$^{\rm 29}$, 
N.~Agrawal\,\orcidlink{0000-0003-0348-9836}\,$^{\rm 25}$, 
Z.~Ahammed\,\orcidlink{0000-0001-5241-7412}\,$^{\rm 132}$, 
S.~Ahmad\,\orcidlink{0000-0003-0497-5705}\,$^{\rm 15}$, 
I.~Ahuja\,\orcidlink{0000-0002-4417-1392}\,$^{\rm 36}$, 
Z.~Akbar$^{\rm 80}$
% ...

\section*{Affiliation Notes}

$^{\rm I}$ Deceased\\
$^{\rm II}$ Also at: Max-Planck-Institut fur Physik, Munich, Germany\\
$^{\rm III}$ Also at: Czech Technical University in Prague (CZ)\\
$^{\rm IV}$ Also at: Italian National Agency for New Technologies, Energy and Sustainable Economic Development (ENEA), Bologna, Italy\\
$^{\rm V}$ Also at: Instituto de Fisica da Universidade de Sao Paulo\\
$^{\rm VI}$ Also at: Dipartimento DET del Politecnico di Torino, Turin, Italy\\
$^{\rm VII}$ Also at: Department of Applied Physics, Aligarh Muslim University, Aligarh, India\\
$^{\rm VIII}$ Also at: Institute of Theoretical Physics, University of Wroclaw, Poland\\
$^{\rm IX}$ Also at: Facultad de Ciencias, Universidad Nacional Aut\'{o}noma de M\'{e}xico, Mexico City, Mexico\\

\section*{Collaboration Institutes}

$^{1}$ A.I. Alikhanyan National Science Laboratory (Yerevan Physics Institute) Foundation, Yerevan, Armenia\\
$^{2}$ AGH University of Krakow, Cracow, Poland\\
$^{3}$ Bogolyubov Institute for Theoretical Physics, National Academy of Sciences of Ukraine, Kyiv, Ukraine\\
$^{4}$ Bose Institute, Department of Physics  and Centre for Astroparticle Physics and Space Science (CAPSS), Kolkata, India\\
$^{5}$ California Polytechnic State University, San Luis Obispo, California, United States\\
$^{6}$ Central China Normal University, Wuhan, China\\
$^{7}$ Centro de Aplicaciones Tecnol\'{o}gicas y Desarrollo Nuclear (CEADEN), Havana, Cuba\\
$^{8}$ Centro de Investigaci\'{o}n y de Estudios Avanzados (CINVESTAV), Mexico City and M\'{e}rida, Mexico\\
$^{9}$ Chicago State University, Chicago, Illinois, United States\\
$^{10}$ China Nuclear Data Center, China Institute of Atomic Energy, Beijing, China\\
$^{11}$ China University of Geosciences, Wuhan, China\\
$^{12}$ Chungbuk National University, Cheongju, Republic of Korea\\
$^{13}$ Comenius University Bratislava, Faculty of Mathematics, Physics and Informatics, Bratislava, Slovak Republic\\
$^{14}$ Creighton University, Omaha, Nebraska, United States\\
$^{15}$ Department of Physics, Aligarh Muslim University, Aligarh, India\\
$^{16}$ Department of Physics, Pusan National University, Pusan, Republic of Korea\\
$^{17}$ Department of Physics, Sejong University, Seoul, Republic of Korea\\
$^{18}$ Department of Physics, University of California, Berkeley, California, United States\\
$^{19}$ Department of Physics, University of Oslo, Oslo, Norway\\
$^{20}$ Department of Physics and Technology, University of Bergen, Bergen, Norway\\
$^{21}$ Dipartimento di Fisica, Universit\`{a} di Pavia, Pavia, Italy\\
$^{22}$ Dipartimento di Fisica dell'Universit\`{a} and Sezione INFN, Cagliari, Italy\\
$^{23}$ Dipartimento di Fisica dell'Universit\`{a} and Sezione INFN, Trieste, Italy\\
$^{24}$ Dipartimento di Fisica dell'Universit\`{a} and Sezione INFN, Turin, Italy\\
$^{25}$ Dipartimento di Fisica e Astronomia dell'Universit\`{a} and Sezione INFN, Bologna, Italy\\
$^{26}$ Dipartimento di Fisica e Astronomia dell'Universit\`{a} and Sezione INFN, Catania, Italy\\
$^{27}$ Dipartimento di Fisica e Astronomia dell'Universit\`{a} and Sezione INFN, Padova, Italy\\
$^{28}$ Dipartimento di Fisica `E.R.~Caianiello' dell'Universit\`{a} and Gruppo Collegato INFN, Salerno, Italy\\
$^{29}$ Dipartimento DISAT del Politecnico and Sezione INFN, Turin, Italy\\
$^{30}$ Dipartimento di Scienze MIFT, Universit\`{a} di Messina, Messina, Italy\\
$^{31}$ Dipartimento Interateneo di Fisica `M.~Merlin' and Sezione INFN, Bari, Italy\\
$^{32}$ European Organization for Nuclear Research (CERN), Geneva, Switzerland\\
$^{33}$ Faculty of Electrical Engineering, Mechanical Engineering and Naval Architecture, University of Split, Split, Croatia\\
$^{34}$ Faculty of Nuclear Sciences and Physical Engineering, Czech Technical University in Prague, Prague, Czech Republic\\
$^{35}$ Faculty of Physics, Sofia University, Sofia, Bulgaria\\
$^{36}$ Faculty of Science, P.J.~\v{S}af\'{a}rik University, Ko\v{s}ice, Slovak Republic\\
$^{37}$ Faculty of Technology, Environmental and Social Sciences, Bergen, Norway\\
$^{38}$ Frankfurt Institute for Advanced Studies, Johann Wolfgang Goethe-Universit\"{a}t Frankfurt, Frankfurt, Germany\\
$^{39}$ Fudan University, Shanghai, China\\
$^{40}$ Gangneung-Wonju National University, Gangneung, Republic of Korea\\
$^{41}$ Gauhati University, Department of Physics, Guwahati, India\\
$^{42}$ Helmholtz-Institut f\"{u}r Strahlen- und Kernphysik, Rheinische Friedrich-Wilhelms-Universit\"{a}t Bonn, Bonn, Germany\\
$^{43}$ Helsinki Institute of Physics (HIP), Helsinki, Finland\\
$^{44}$ High Energy Physics Group,  Universidad Aut\'{o}noma de Puebla, Puebla, Mexico\\
$^{45}$ Horia Hulubei National Institute of Physics and Nuclear Engineering, Bucharest, Romania\\
$^{46}$ HUN-REN Wigner Research Centre for Physics, Budapest, Hungary\\
$^{47}$ Indian Institute of Technology Bombay (IIT), Mumbai, India\\
$^{48}$ Indian Institute of Technology Indore, Indore, India\\
$^{49}$ INFN, Laboratori Nazionali di Frascati, Frascati, Italy\\
$^{50}$ INFN, Sezione di Bari, Bari, Italy\\
$^{51}$ INFN, Sezione di Bologna, Bologna, Italy\\
$^{52}$ INFN, Sezione di Cagliari, Cagliari, Italy\\
$^{53}$ INFN, Sezione di Catania, Catania, Italy\\
$^{54}$ INFN, Sezione di Padova, Padova, Italy\\
$^{55}$ INFN, Sezione di Pavia, Pavia, Italy\\
$^{56}$ INFN, Sezione di Torino, Turin, Italy\\
$^{57}$ INFN, Sezione di Trieste, Trieste, Italy\\
$^{58}$ Inha University, Incheon, Republic of Korea\\
$^{59}$ Institute for Gravitational and Subatomic Physics (GRASP), Utrecht University/Nikhef, Utrecht, Netherlands\\
$^{60}$ Institute of Experimental Physics, Slovak Academy of Sciences, Ko\v{s}ice, Slovak Republic\\
$^{61}$ Institute of Physics, Homi Bhabha National Institute, Bhubaneswar, India\\
$^{62}$ Institute of Physics of the Czech Academy of Sciences, Prague, Czech Republic\\
$^{63}$ Institute of Space Science (ISS), Bucharest, Romania\\
$^{64}$ Institut f\"{u}r Kernphysik, Johann Wolfgang Goethe-Universit\"{a}t Frankfurt, Frankfurt, Germany\\
$^{65}$ Instituto de Ciencias Nucleares, Universidad Nacional Aut\'{o}noma de M\'{e}xico, Mexico City, Mexico\\
$^{66}$ Instituto de F\'{i}sica, Universidade Federal do Rio Grande do Sul (UFRGS), Porto Alegre, Brazil\\
$^{67}$ Instituto de F\'{\i}sica, Universidad Nacional Aut\'{o}noma de M\'{e}xico, Mexico City, Mexico\\
$^{68}$ iThemba LABS, National Research Foundation, Somerset West, South Africa\\
$^{69}$ Jeonbuk National University, Jeonju, Republic of Korea\\
$^{70}$ Johann-Wolfgang-Goethe Universit\"{a}t Frankfurt Institut f\"{u}r Informatik, Fachbereich Informatik und Mathematik, Frankfurt, Germany\\
$^{71}$ Laboratoire de Physique Subatomique et de Cosmologie, Universit\'{e} Grenoble-Alpes, CNRS-IN2P3, Grenoble, France\\
$^{72}$ Lawrence Berkeley National Laboratory, Berkeley, California, United States\\
$^{73}$ Lund University Department of Physics, Division of Particle Physics, Lund, Sweden\\
$^{74}$ Nagasaki Institute of Applied Science, Nagasaki, Japan\\
$^{75}$ Nara Women{'}s University (NWU), Nara, Japan\\
$^{76}$ National and Kapodistrian University of Athens, School of Science, Department of Physics , Athens, Greece\\
$^{77}$ National Centre for Nuclear Research, Warsaw, Poland\\
$^{78}$ National Institute of Science Education and Research, Homi Bhabha National Institute, Jatni, India\\
$^{79}$ National Nuclear Research Center, Baku, Azerbaijan\\
$^{80}$ National Research and Innovation Agency - BRIN, Jakarta, Indonesia\\
$^{81}$ Niels Bohr Institute, University of Copenhagen, Copenhagen, Denmark\\
$^{82}$ Nikhef, National institute for subatomic physics, Amsterdam, Netherlands\\
$^{83}$ Nuclear Physics Group, STFC Daresbury Laboratory, Daresbury, United Kingdom\\
$^{84}$ Nuclear Physics Institute of the Czech Academy of Sciences, Husinec-\v{R}e\v{z}, Czech Republic\\
$^{85}$ Oak Ridge National Laboratory, Oak Ridge, Tennessee, United States\\
$^{86}$ Ohio State University, Columbus, Ohio, United States\\
$^{87}$ Physics department, Faculty of science, University of Zagreb, Zagreb, Croatia\\
$^{88}$ Physics Department, Panjab University, Chandigarh, India\\
$^{89}$ Physics Department, University of Jammu, Jammu, India\\
$^{90}$ Physics Program and International Institute for Sustainability with Knotted Chiral Meta Matter (WPI-SKCM$^{2}$), Hiroshima University, Hiroshima, Japan\\
$^{91}$ Physikalisches Institut, Eberhard-Karls-Universit\"{a}t T\"{u}bingen, T\"{u}bingen, Germany\\
$^{92}$ Physikalisches Institut, Ruprecht-Karls-Universit\"{a}t Heidelberg, Heidelberg, Germany\\
$^{93}$ Physik Department, Technische Universit\"{a}t M\"{u}nchen, Munich, Germany\\
$^{94}$ Politecnico di Bari and Sezione INFN, Bari, Italy\\
$^{95}$ Research Division and ExtreMe Matter Institute EMMI, GSI Helmholtzzentrum f\"ur Schwerionenforschung GmbH, Darmstadt, Germany\\
$^{96}$ Saga University, Saga, Japan\\
$^{97}$ Saha Institute of Nuclear Physics, Homi Bhabha National Institute, Kolkata, India\\
$^{98}$ School of Physics and Astronomy, University of Birmingham, Birmingham, United Kingdom\\
$^{99}$ Secci\'{o}n F\'{\i}sica, Departamento de Ciencias, Pontificia Universidad Cat\'{o}lica del Per\'{u}, Lima, Peru\\
$^{100}$ Stefan Meyer Institut f\"{u}r Subatomare Physik (SMI), Vienna, Austria\\
$^{101}$ SUBATECH, IMT Atlantique, Nantes Universit\'{e}, CNRS-IN2P3, Nantes, France\\
$^{102}$ Sungkyunkwan University, Suwon City, Republic of Korea\\
$^{103}$ Suranaree University of Technology, Nakhon Ratchasima, Thailand\\
$^{104}$ Technical University of Ko\v{s}ice, Ko\v{s}ice, Slovak Republic\\
$^{105}$ The Henryk Niewodniczanski Institute of Nuclear Physics, Polish Academy of Sciences, Cracow, Poland\\
$^{106}$ The University of Texas at Austin, Austin, Texas, United States\\
$^{107}$ Universidad Aut\'{o}noma de Sinaloa, Culiac\'{a}n, Mexico\\
$^{108}$ Universidade de S\~{a}o Paulo (USP), S\~{a}o Paulo, Brazil\\
$^{109}$ Universidade Estadual de Campinas (UNICAMP), Campinas, Brazil\\
$^{110}$ Universidade Federal do ABC, Santo Andre, Brazil\\
$^{111}$ Universitatea Nationala de Stiinta si Tehnologie Politehnica Bucuresti, Bucharest, Romania\\
$^{112}$ University of Derby, Derby, United Kingdom\\
$^{113}$ University of Houston, Houston, Texas, United States\\
$^{114}$ University of Jyv\"{a}skyl\"{a}, Jyv\"{a}skyl\"{a}, Finland\\
$^{115}$ University of Kansas, Lawrence, Kansas, United States\\
$^{116}$ University of Liverpool, Liverpool, United Kingdom\\
$^{117}$ University of Science and Technology of China, Hefei, China\\
$^{118}$ University of South-Eastern Norway, Kongsberg, Norway\\
$^{119}$ University of Tennessee, Knoxville, Tennessee, United States\\
$^{120}$ University of the Witwatersrand, Johannesburg, South Africa\\
$^{121}$ University of Tokyo, Tokyo, Japan\\
$^{122}$ University of Tsukuba, Tsukuba, Japan\\
$^{123}$ Universit\"{a}t M\"{u}nster, Institut f\"{u}r Kernphysik, M\"{u}nster, Germany\\
$^{124}$ Universit\'{e} Clermont Auvergne, CNRS/IN2P3, LPC, Clermont-Ferrand, France\\
$^{125}$ Universit\'{e} de Lyon, CNRS/IN2P3, Institut de Physique des 2 Infinis de Lyon, Lyon, France\\
$^{126}$ Universit\'{e} de Strasbourg, CNRS, IPHC UMR 7178, F-67000 Strasbourg, France, Strasbourg, France\\
$^{127}$ Universit\'{e} Paris-Saclay, Centre d'Etudes de Saclay (CEA), IRFU, D\'{e}partment de Physique Nucl\'{e}aire (DPhN), Saclay, France\\
$^{128}$ Universit\'{e}  Paris-Saclay, CNRS/IN2P3, IJCLab, Orsay, France\\
$^{129}$ Universit\`{a} degli Studi di Foggia, Foggia, Italy\\
$^{130}$ Universit\`{a} del Piemonte Orientale, Vercelli, Italy\\
$^{131}$ Universit\`{a} di Brescia, Brescia, Italy\\
$^{132}$ Variable Energy Cyclotron Centre, Homi Bhabha National Institute, Kolkata, India\\
$^{133}$ Warsaw University of Technology, Warsaw, Poland\\
$^{134}$ Wayne State University, Detroit, Michigan, United States\\
$^{135}$ Yale University, New Haven, Connecticut, United States\\
$^{136}$ Yildiz Technical University, Istanbul, Turkey\\
$^{137}$ Yonsei University, Seoul, Republic of Korea\\
$^{138}$ Affiliated with an institute formerly covered by a cooperation agreement with CERN\\
$^{139}$ Affiliated with an international laboratory covered by a cooperation agreement with CERN.\\

\end{flushleft}